\documentclass[cits]{PoS}

\usepackage[authoryear]{natbib}
\bibpunct{(}{)}{;}{a}{,}{,}
\usepackage{graphicx}

\usepackage{amssymb}
\usepackage{rotating}

\title{Dark Matter Annihilation in the light of EGRET, HEAT, WMAP, INTEGRAL and ROSAT}

\ShortTitle{DMA in the light of EGRET, HEAT, WMAP, INTEGRAL and ROSAT}

\author{\speaker{Iris Gebauer}\\
        Institut f\"{u}r Experimentelle Kernphysik, Universi\"{a}t Karlsruhe\\
        E-mail: \email{gebauer@ekp.uni-karlsruhe.de}}

\abstract{The ROSAT Galactic wind observations confirm that our Galaxy
launches supernova (SN) driven Galactic winds with wind speeds of about
150 km/s in
the Galactic plane. Galactic winds of this strength are incompatible
with current isotropic models for Cosmic Ray (CR) transport as implemented in the GALPROP code.\\
In order to reproduce our local CRs in the presence of Galactic winds, charged CRs are required to be much more localized than in the
standard isotropic GALPROP models. This requires that anisotropic
diffusion is the dominant diffusion mode in the interstellar medium (ISM), particularly that the
diffusion in the disk and in the halo are different. In addition small
scale phenomena such as trapping by molecular cloud complexes and the
structure of our local environment (the local bubble or the local fluff) might influence the secondary CR production rate and our
local CR density gradients.\\
We introduce an anisotropic convection driven transport model (aCDM)
which is consistent with the Galactic wind observations by ROSAT. This
automatically explains the large bulge/disk ratio as observed by
INTEGRAL. Furthermore such models predict an increase in the  $e^+/(e^++e^-)$-fraction as
observed by PAMELA and HEAT, if the synchrotron constraints in the 408 MHz and
WMAP range are taken into account. This increase originates entirely from the
transport properties of electrons and positrons, no additional
contribution from Dark Matter (DM) is required. The aCDM is able to explain the absence of a positron annihilation signal from molecular clouds
(MCs) as observed by INTEGRAL by virtue of a mechanism which confines
and isotropizes CRs between MCs.
Unlike isotropic models, this model does not rely on a flattened source distribution or a strong
increase in the $X_{CO}$ scaling factor, but uses the supernova remnant (SNR)
distribution as source distribution for CRs and a flat $X_{CO}$ scaling
factor. We find that the EGRET excess of diffuse $\gamma$-rays currently cannot be
explained by astrophysical effects in this type of model and that
the  interpretation of the EGRET excess as Dark Matter annihilation (DMA) is
perfectly consistent with all observational constraints from local CR
fluxes and synchrotron radiation.}

\FullConference{Identification of dark matter 2008\\
		 August 18-22, 2008\\
		 Stockholm, Sweden}

\begin{document}

\section{Introduction}
Current CR transport models as implemented in the GALPROP
code\footnote{The GALPROP code is publicly available from http://galprop.stanford.edu} are able
to either explain our local charged CRs ({\it conventional}
models) or the diffuse Galactic $\gamma$-rays ({\it optimized}
models). The reason for this is that the locally observed proton spectrum is incompatible with
the observed $\gamma$-rays by both, spectral shape and absolute
flux. In isotropic diffusion dominated transport mo\-dels this
inconsistency cannot be explained by transport effects, because our
local CRs are strongly linked to the interstellar CR population. As a
possible explanation an additional component from DMA has been proposed \cite{us}.
However, in this case isotropic transport models predict too many local
antiprotons \cite{bergstrom1}. It is shown that anisotropic diffusion
models are constistent with the DMA interpretation of the EGRET excess.\\

\section{Anisotropic Convection Driven Transport Models}
\begin{figure}
\includegraphics[width=0.32\textwidth]{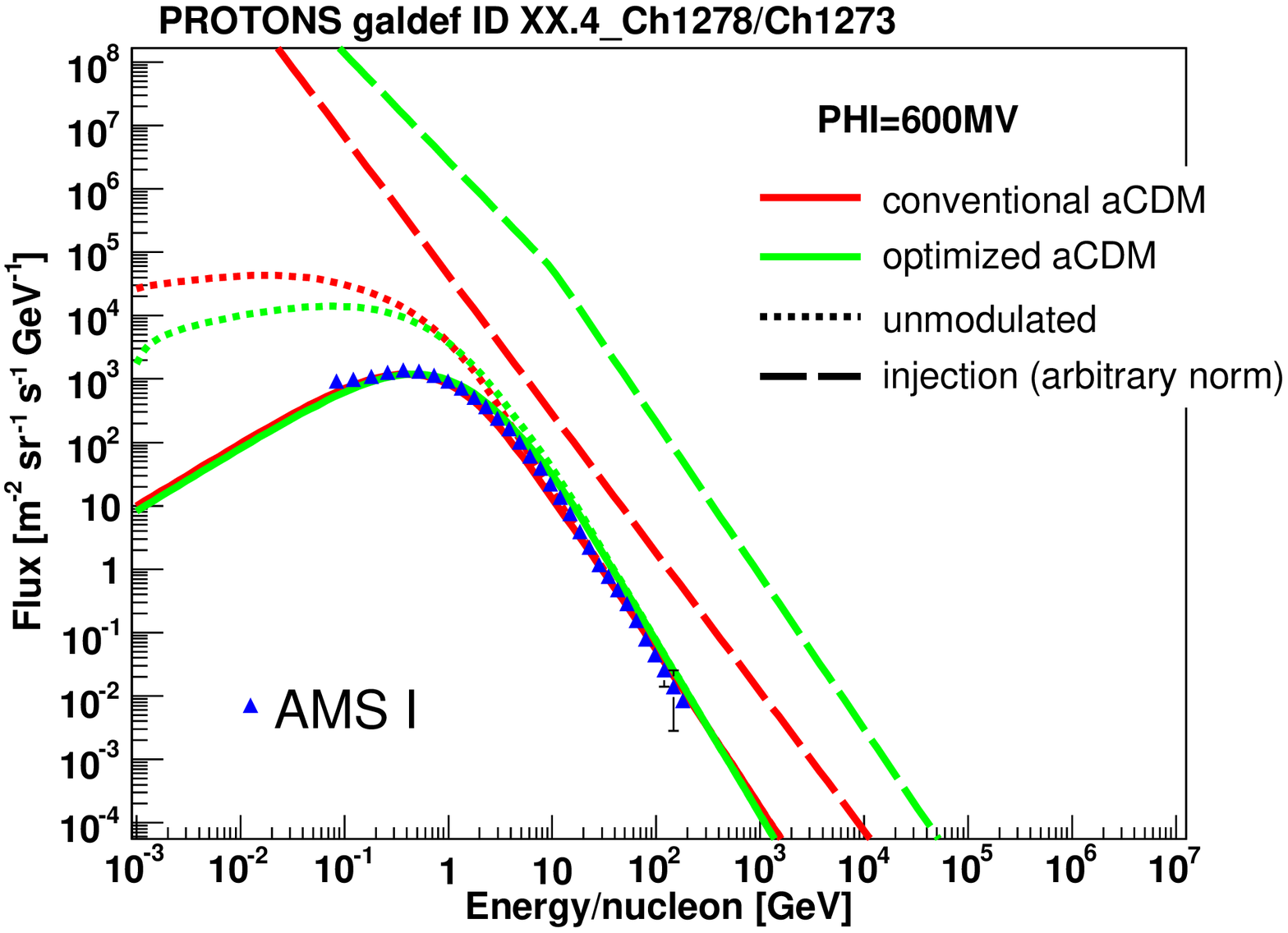}
\includegraphics[width=0.32\textwidth]{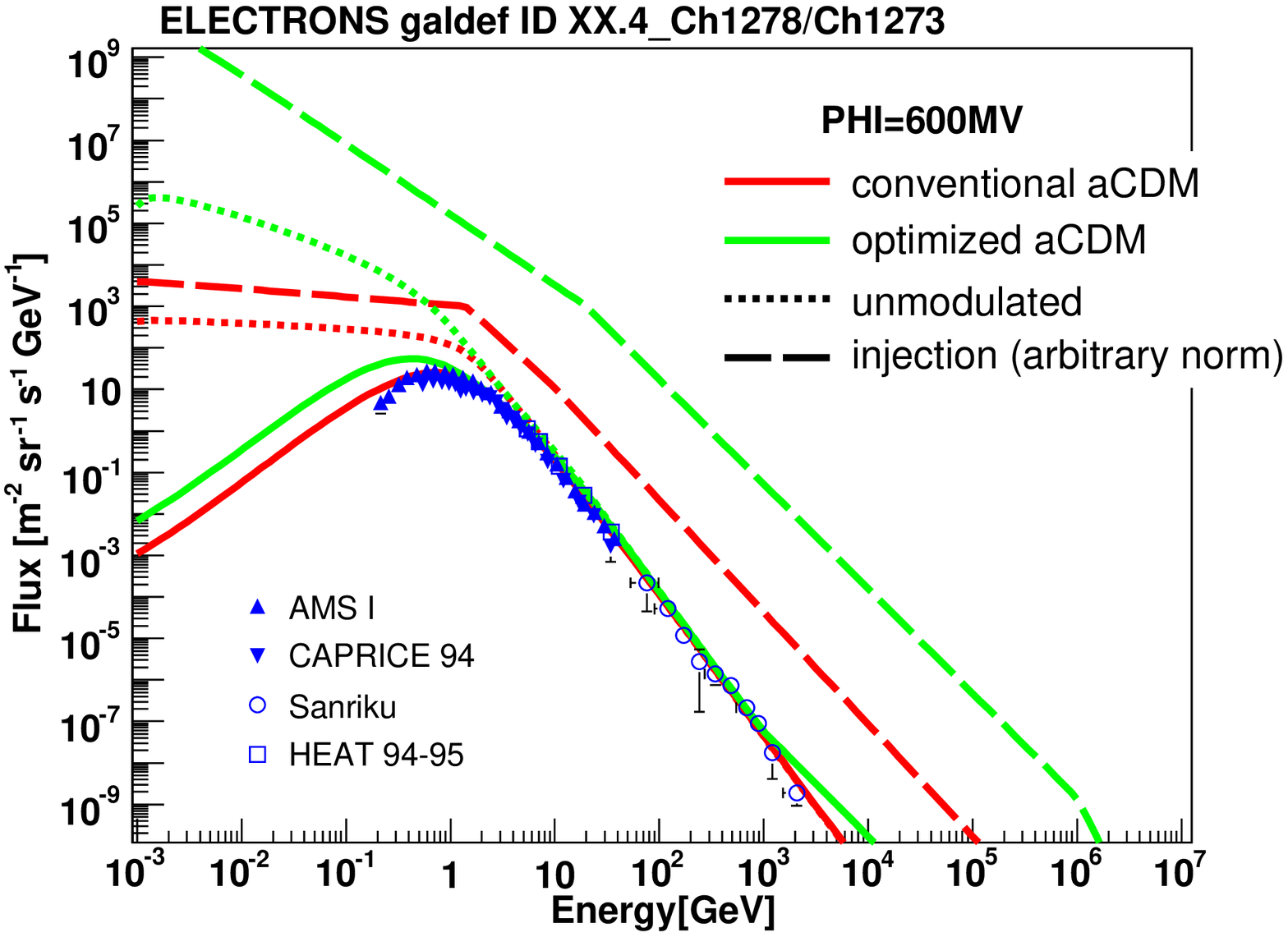}
\includegraphics[width=0.32\textwidth]{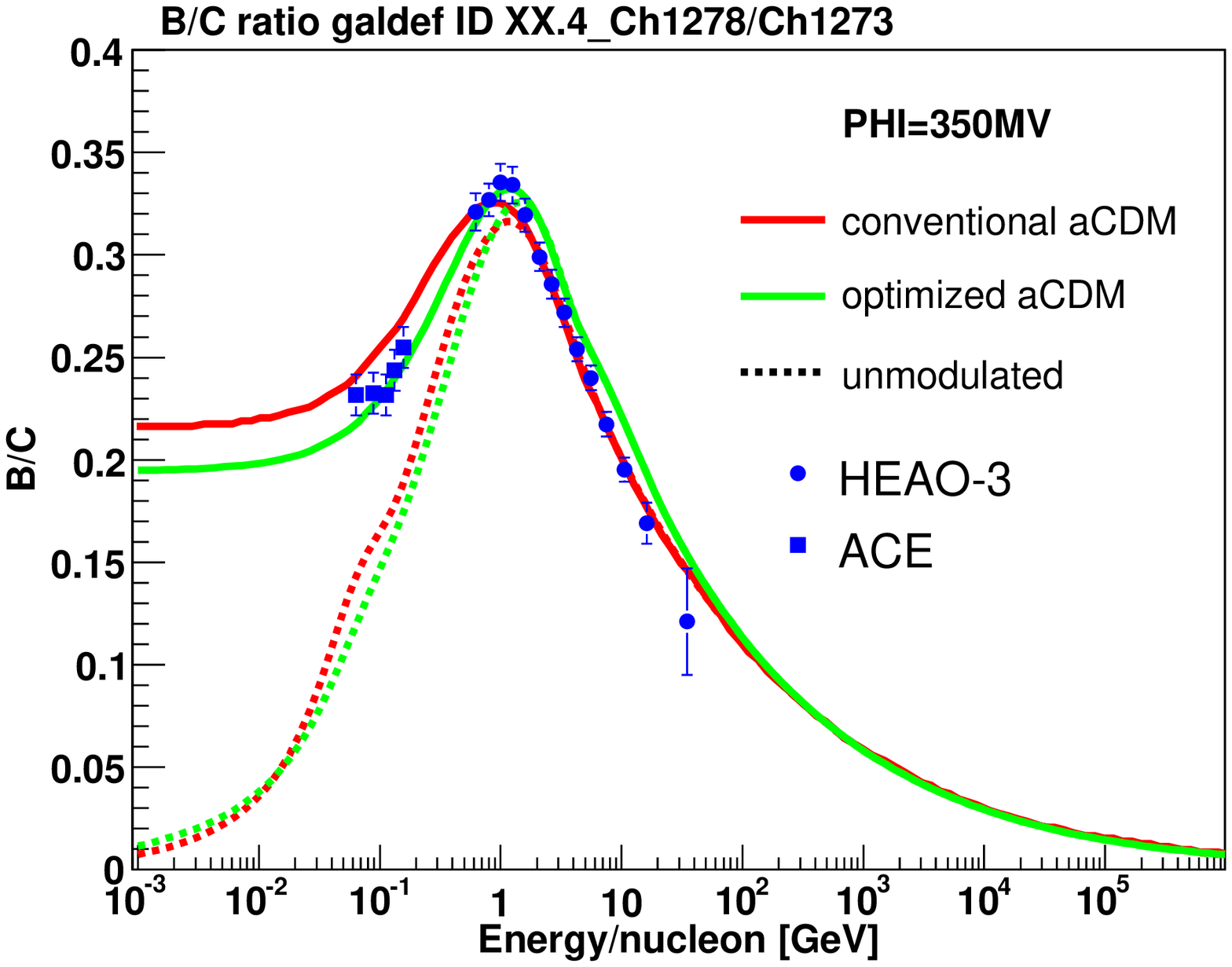}
\caption{{\it Left:} Local proton flux for a conventional (red) and optimized
 aCDM (green). The full lines correspond to local modulated fluxes,
 the dotted lines are local unmodulated spectra and the dashed lines
 are the injection spectra. {\it Center:}
 Local electron spectrum, line coding as for protons. {\it Right:} Local
 $B/C$ ratio in an aCDM, line coding as for protons.}\label{f1}
\end{figure}

\begin{figure}

\includegraphics[width=0.32\textwidth]{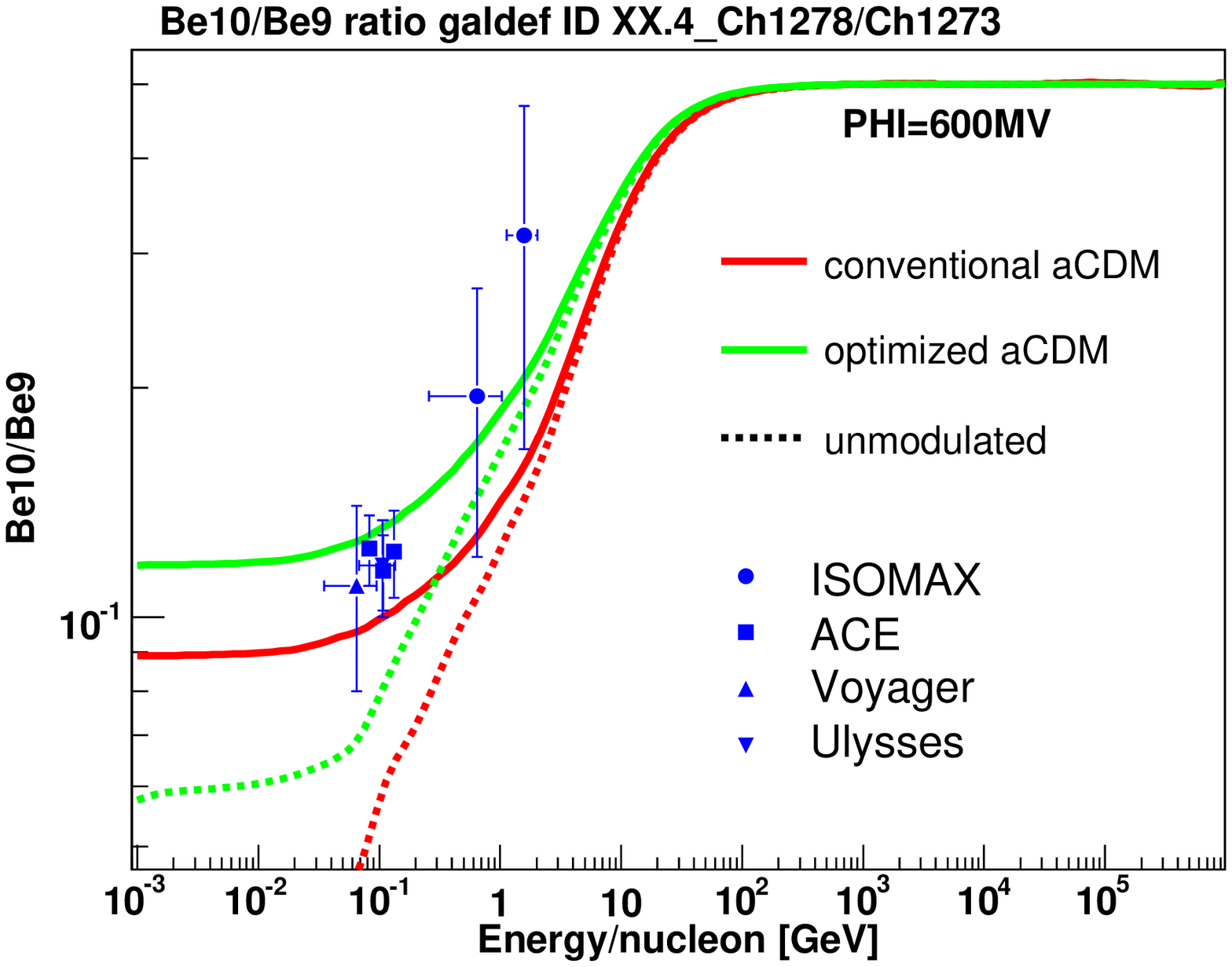}
\includegraphics[width=0.32\textwidth]{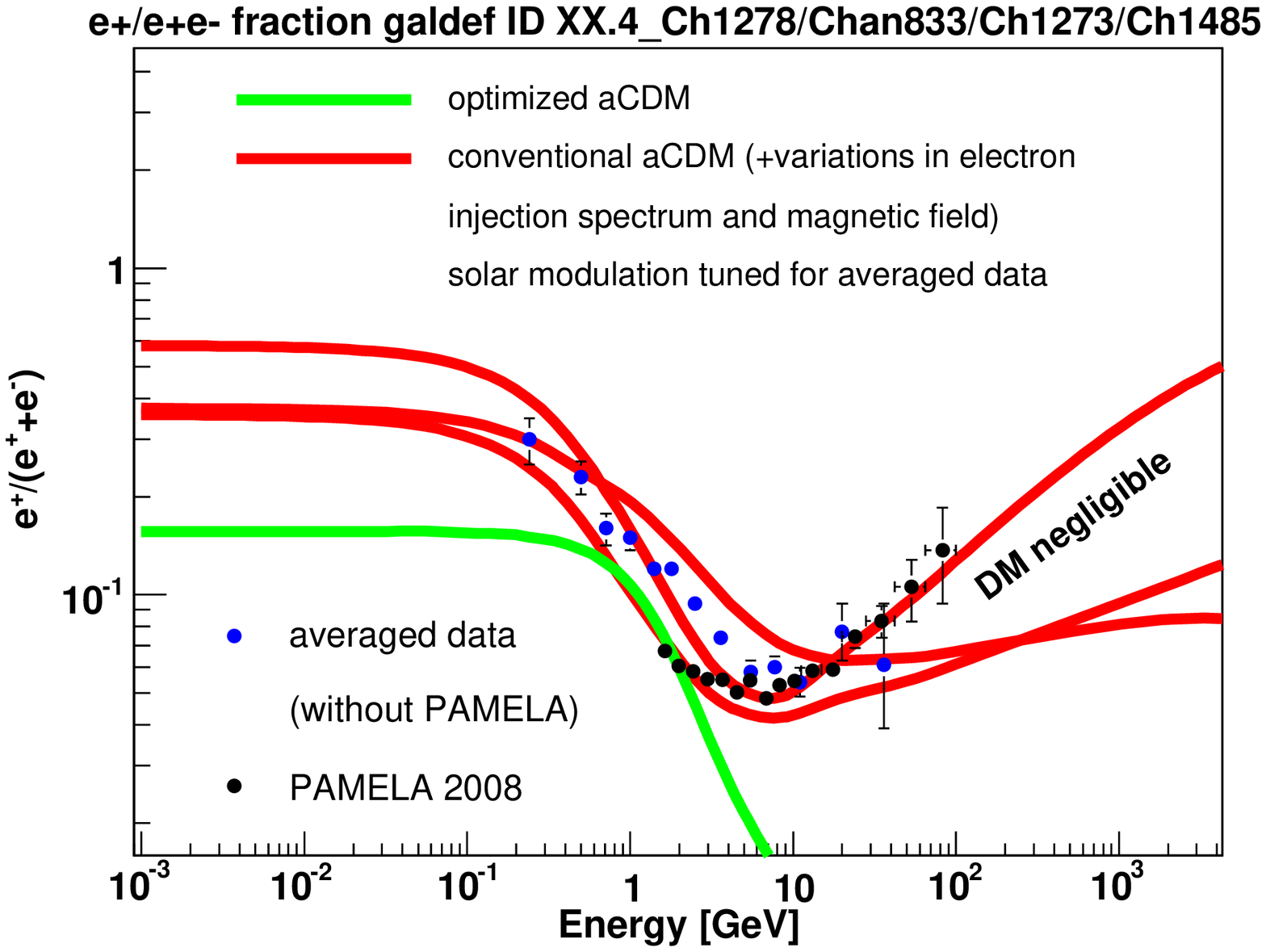}
\includegraphics[width=0.32\textwidth]{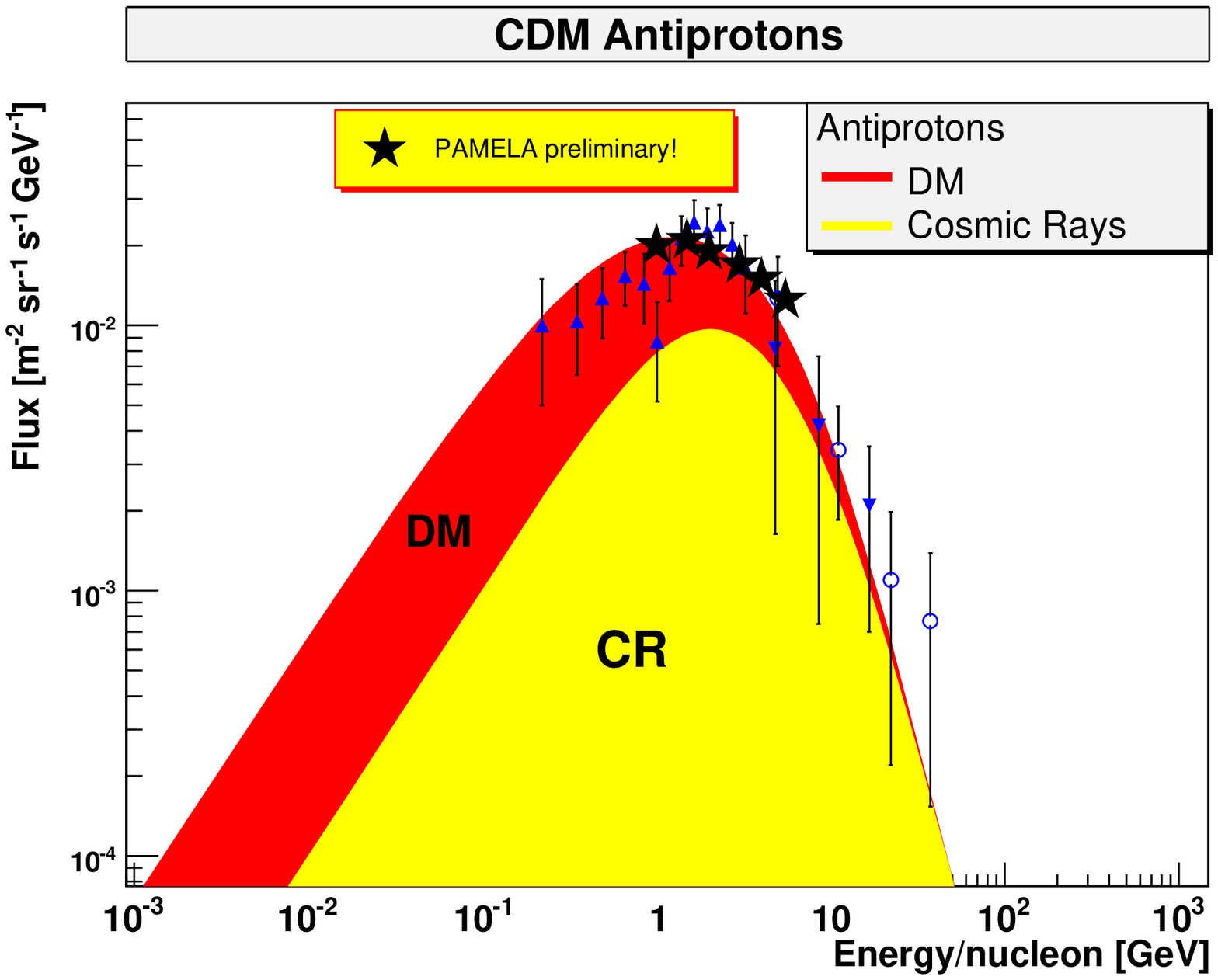}
\caption{ {\it Left:} Local $^{10}Be/^9Be$-fraction for an aCDM, line 
 coding as for protons in figure 1.
{\it Center:} Local
 $e^+/e^++e^-$-fraction for an optimized aCDM (green) and a conventional
 aCDM (red). For the conventional aCDM the different red lines
 correspond to different combinations of electron injection spectra and
 magnetic fields, which are correlated by a fit to the synchrotron
 radiation spectra of the Haslam skymap and WMAP data. The averaged data
 are taken from \cite{olzem} and do not include the PAMELA results. {\it Right:} Local antiprotons in a
 conventional aCDM. The contribution from CRs is very similar to the
 predictions of the isotropic models, while the contribution from DMA is
 reduced by virtue of the smaller collection volume.
}\label{f2}
\end{figure}

\begin{figure}
\includegraphics[width=0.32\textwidth]{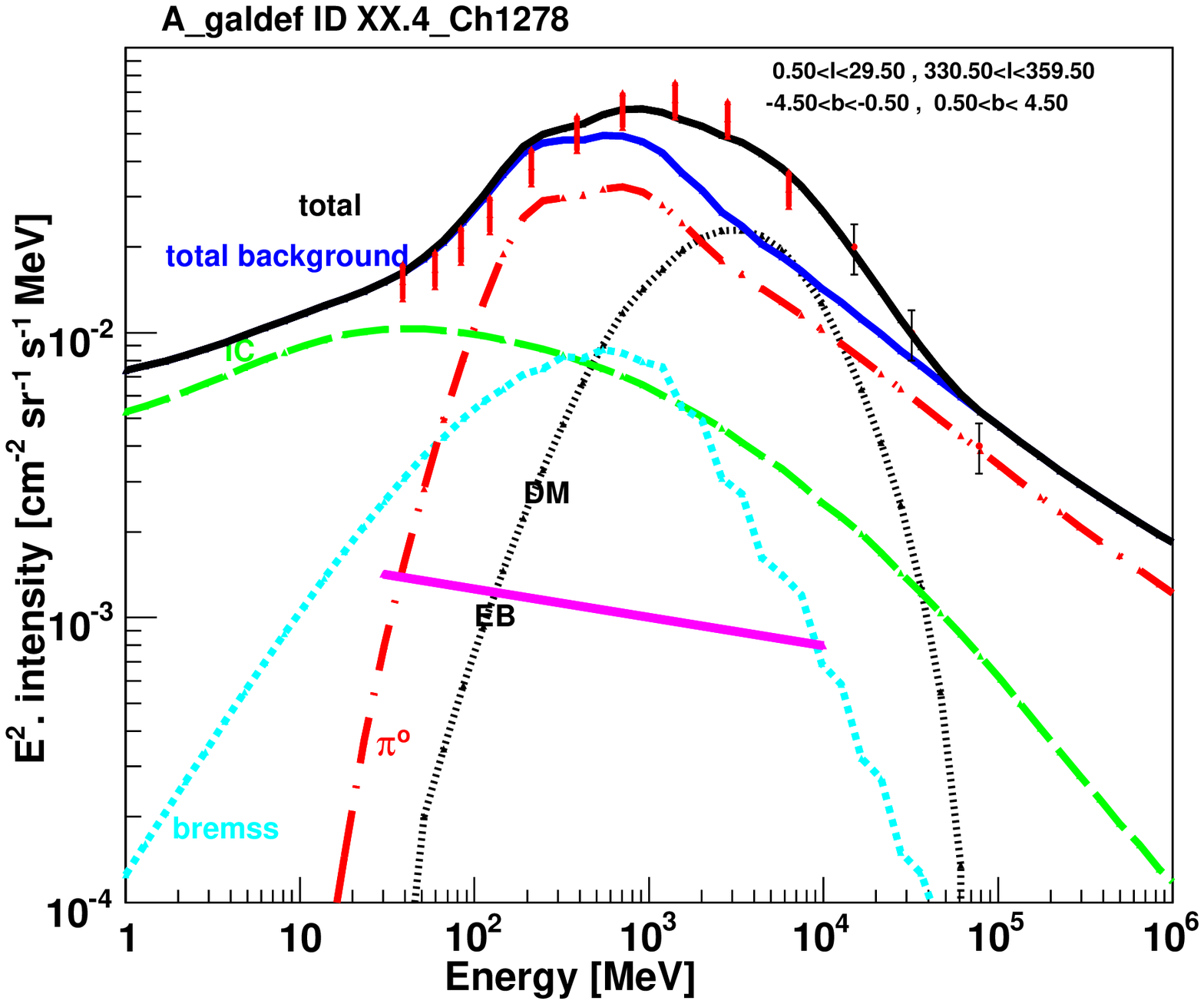}
\includegraphics[width=0.32\textwidth]{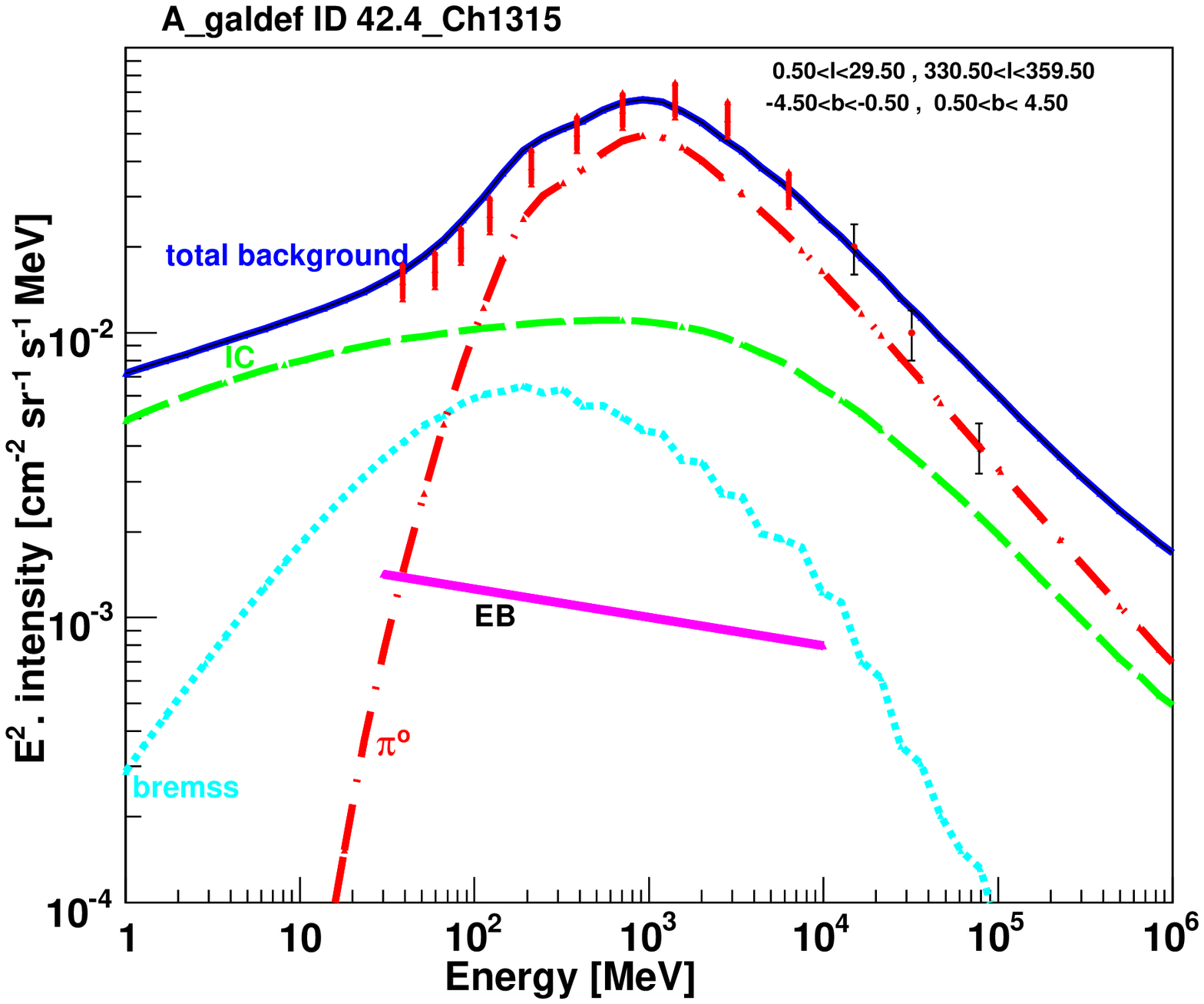}
\includegraphics[width=0.32\textwidth]{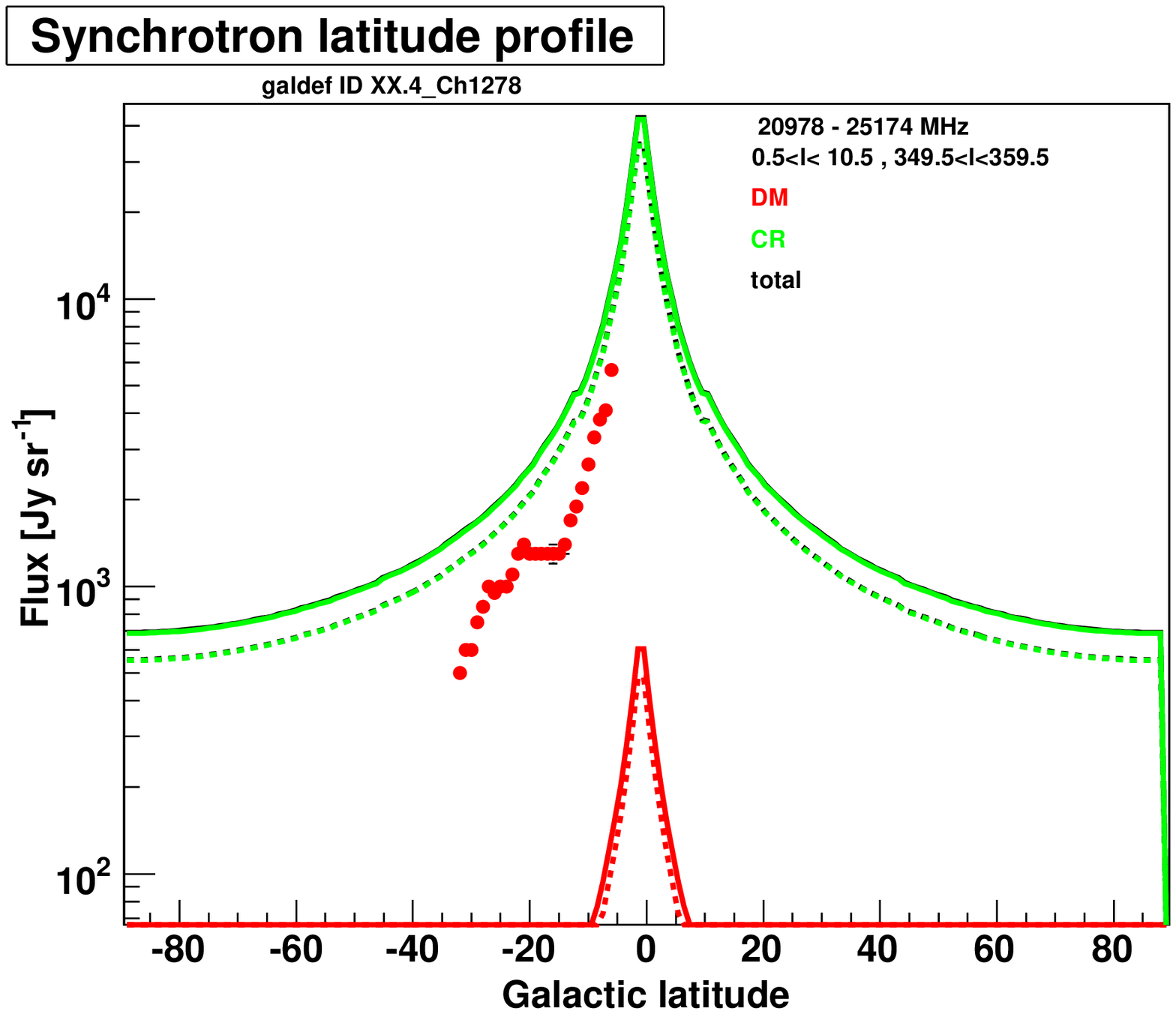}
\caption{ Diffuse $\gamma$-rays from the Galactic center
 region in a conventional ({\it left}) and optimized ({\it center})
 aCDM. {\it Right:} Latitude distribution of the synchrotron
 radiation in the small longitude range where the haze has been
 measured. The top solid curve is the total flux, the red dots represent
 the haze and the lower curve is the contribution from the DMA
 interpretation of the EGRET excess.
}\label{f3}
\end{figure}

The EGRET excess is not the only challenge for Galactic transport
models. Self-consistent Galactic wind calculations have led to the
conclusion that convective transport plays a non-negligible r\^{o}le in
our Galaxy. A recent analysis of the
ROSAT X-ray data has confirmed that our Galaxy indeed launches winds with
speeds of 150 km/s in the plane at the position of the sun and up to
800 km/s in
the halo  \citep{breitschwerdt_nature}. Although this is a comparably
low speed (starburst Galaxies launch winds with speeds up to 3000
km/s), the impact upon CR transport is significant, because convective
transport occurs only in the direction perpendicular to the plane. As expected, the wind
velocities appear to be roughly proportional to the SNR distribution. Transport models with isotropic diffusion can allow for convection
velocity gradients of only 10 km/s/kpc (with zero wind speed at z=0 kpc), since for larger
velocities the constraints from radioactive isotopes and secondary
particle production cannot be met \cite{strong_rev}. 
To account for the observed wind velocities anisotropic
diffusion modes have to be considered.
Convection velocities compatible with the ROSAT observations
significantly reduce the distance CRs travel in our Galaxy before they
enter the convection zone and escape, thus leading to a model which
allows for both - an effective decoupling of local CRs from the global
CR population as required by an astrophysical explanation of the EGRET
excess and a small collection volume for all CR species including
antiprotons from DMA.
Such a model not only decouples the local CR spectra from the Galactic
average, it also explains the large bulge/disk ratio as observed in the
511 keV line by INTEGRAL \citep{integral} purely by the properties of CR
transport. The
basic observation is the high intensity from the region of the Galactic
bulge and a rather low signal from the disk, although the opposite is
expected \cite{prantzos}. The annihilation signal in the disk can  be entirely explained by
the decay of $^{26}Al$ from core collapse SNs, leaving no room for positrons from SNIa. Isotropic transport models have to invoke additional
sources
which are confined to the bulge in order to
partially \cite{integral_nature} or entirely \cite{boehm} explain the observed signal from the bulge, but they cannot explain  
why there is no annihilation signal from SNIa in the disk.
In an aCDM positrons from SNIa are produced predominantly in regions
with strong Galactic winds and so the MeV
positrons from SNIa are almost immediately blown into the halo. In the bulge, where the lower SN rate and
the strong gravitational potential reduce the wind strength,
positrons have enough time to annihilate. A second surprising result from INTEGRAL is the
observation that positrons do not annihilate in molecular clouds (MCs). A
spectral analysis of the annihilation line indicates that practically
all annihilation takes place in the atomic component of the gas,
although molecular clouds have an average density of more than half of
the gas \cite{jean}. One could argue that the volume filling factor of molecular
clouds is too small to be found by positrons. However, the high magnetic
field inside MCs appears to correlate with the interstellar magnetic
field \citet{han}. In such a model MCs will act as magnetic mirrors and
positrons are likely to be reflected and confined between MCs
\cite{chandran}. \\
Note, that Galactic winds in agreement with ROSAT are also able to
explain the observed soft gradient in diffuse $\gamma$-rays without
invoking a flattened source distribution or an increase in the
$X_{CO}$-scaling factor: the peak in the gas distribution coincides
with the peak in the SNR distribution, consequently the CR interaction
rate in this region, which leads to a relative overproduction of diffuse
$\gamma$-rays in the isotropic models, is reduced in an aCDM.\\
INTEGRAL and ROSAT form strong constraints for any transport model.
We have implemented the transport picture described above into the
publicly available GALPROP code. The main features of our implementation
are: i) fast propagation perpendicular to the disk by turbulent
diffusion and convection in agreement with the ROSAT X-ray data ii) slow
diffusion in the disk and a possible trapping mechanism by molecular
clouds which would increase the CR interaction rate in the atomic
component of the gas. In addition we allow for local variations in all
transport parameters in order to account for our local Galactic
environment. Compared to isotropic
models CRs travel smaller distances in an aCDM leading to a stronger
impact of local transport phenomena.\\
We tune the transport parameters for local CR
spectra and allow for an additional contribution from DMA in order to
describe both, the diffuse $\gamma$-rays and local CR spectra ({\it conventional} aCDM). Local protons, electrons, $B/C$-fraction,
$^{10}Be/^9Be$-fraction, $e^+/(e^++e^-)$-fraction, diffuse $\gamma$-rays
and the 22 GHz synchrotron profile for the haze region are shown in figures
\ref{f1} to \ref{f3}. A breakless proton injection spectrum and a
breakless diffusion coefficient turn out
to be sufficient to describe the locally observed CR fluxes and the
diffuse $\gamma$-rays. Note, that
in this model the increase in  $e^+/(e^++e^-)$-fraction, as observed by
PAMELA, is a natural consequence of CR transport: the breaks in
$e^+/(e^++e^-)$-fraction are mainly determined by the electron and
proton injection spectra, momentum losses and gains only play a minor
r\^ole. In a conventional aCDM this naturally yields a positron spectrum which is
slightly harder than the electron spectrum and consequently gives rise to
the observed increase in $e^+/(e^++e^-)$-fraction (this interpretation is
strongly supported by the proton and electron measuments from AMS up to 30
GeV, which show an even steeper increase in the $p/(p+e^-)$-fraction).   
Compared to isotropic models the contribution from DMA in local CRs is small, since
most of the positrons and antiprotons from DMA are produced in the halo, thus drifting
away by convection. 
Note, that in an aCDM no additional fine-tuning is
required to describe the data on local antiprotons: convection
speeds are fixed by the ROSAT observations and transport parameters
are fixed by local CR spectra and constraints from synchrotron
radiation.
The right side of figure \ref{f3} shows the latitude profile of the
synchrotron radiation in the so-called WMAP haze region for 22GHz.
The WMAP haze consists of an excess of microwave emission from a small
region close to the Galactic center.
 It has been suggested that this signal could be synchrotron emission from relativistic electrons and
positrons, possibly originating from DMA in a cuspy halo \cite{haze_my}.
However, even in a cored profile the synchrotron radiation from
the disk shows a  steep increase as shown in figure \ref{f3}, while
the intensity of synchrotron radiation from DMA (as required to explain
the EGRET excess) is much lower.
In order to check for astrophysical sources of the EGRET excess we tune
the transport parameters with respect to diffuse $\gamma$-rays and try
to reproduce our local charged CRs by variations of our local transport
parameters ({\it optimized} aCDM). Figures \ref{f1} to \ref{f3} show
the local protons, electrons, $B/C$-fraction, $^{10}Be/^9Be$-fraction,
$e^+/(e^++e^-)$-fraction and diffuse $\gamma$-rays in an optimized
aCDM. As in the isotropic model the $\gamma$-rays require a
break in the proton injection spectrum at 10 GeV which is not expected
from the acceleration of protons by SN explosions \cite{berezinsky}.
Unlike the isotropic models, in this type of model we are able to describe the 
observed spectra of protons and nuclei reasonably well by choosing our
local transport parameters accordingly.
Still, in an
optimized aCDM local electrons seem to be incompatible with the
locally observed proton spectrum as can be seen from the
$e^+/(e^++e^-)$-fraction in figure \ref{f2}.

\section{Conclusion}
We have presented a transport model for Galactic CRs compatible with the
wind velocities observed by ROSAT. In such a model the INTEGRAL 511 keV
line and its spectral morphology, the synchrotron radiation observed by WMAP and the locally observed
positron and electron spectra are explained entirely as a consequence of CR
transport. However, without DMA this type of model is currently not able to explain
the diffuse $\gamma$-rays as observed by EGRET by
astrophysical effects in concordance with the local CR spectra. We
showed that the DMA interpretation of the EGRET excess is compatible
with the constraints from CR transport.

\end{document}